\documentclass[11pt]{article}
\usepackage[left=1.25in,top=1in,right=1.25in,bottom=1in]{geometry}
\usepackage{amsmath}
\usepackage{amssymb}
\usepackage{amsbsy}
\usepackage{amsthm}
\usepackage{epsfig}
\usepackage[round]{natbib}
\usepackage{clrscode}
\usepackage{setspace}
\usepackage{multirow}
\usepackage{color}
\usepackage{url}
\usepackage{booktabs}

\usepackage{rotating}

\newcommand{\N}{\mbox{N}}

\newcommand{\E}{\mbox{E}}
\newcommand{\var}{\mbox{var}}

\newcommand{\balpha}{\boldsymbol{\alpha}}

\newcommand{\beps}{\boldsymbol{\epsilon}}
\newcommand{\bxi}{\boldsymbol{\xi}}
\newcommand{\bmu}{\boldsymbol{\mu}}
\newcommand{\boldeta}{\boldsymbol{\eta}}
\newcommand{\bpsi}{\boldsymbol{\psi}}

\newcommand{\bu}{\mathbf{u}}
\newcommand{\be}{\mathbf{e}}
\newcommand{\bv}{\mathbf{v}}
\newcommand{\by}{\mathbf{y}}

\newcommand{\beff}{\mathbf{f}}

\title{An empirical test for Eurozone contagion using an asset-pricing model with heavy-tailed stochastic volatility}

\author{
\textsc{Nicholas G.~Polson} \\
\small{\textit{Booth School of Business}} \\ \small{\textit{University of Chicago}} \\
\\
\textsc{James G.~Scott} \\
\small{\textit{McCombs School of Business}} \\ \small{\textit{University of Texas at Austin}} \\
}

\begin{document}

\maketitle
\begin{abstract}

This paper proposes an empirical test of financial contagion in European equity markets during the tumultuous period of 2008--2011.  Our analysis shows that traditional GARCH and Gaussian stochastic-volatility models are unable to explain two key stylized features of global markets during presumptive contagion periods: shocks to aggregate market volatility can be sudden and explosive, and they are associated with specific directional biases in the cross-section of country-level returns.  Our model repairs this deficit by assuming that the random shocks to volatility are heavy-tailed and correlated cross-sectionally, both with each other and with returns.  The fundamental conclusion of our analysis is that great care is needed in modeling volatility if one wishes to characterize the relationship between volatility and contagion that is predicted by economic theory.

In analyzing daily data, we find evidence for significant contagion effects during the major EU crisis periods of May 2010 and August 2011, where contagion is defined as excess correlation in the residuals from a factor model incorporating global and regional market risk factors.  Some of this excess correlation can be explained by quantifying the impact of shocks to aggregate volatility in the cross-section of expected returns---but only, it turns out, if one is extremely careful in accounting for the explosive nature of these shocks.  We show that global markets have time-varying cross-sectional sensitivities to these shocks, and that high sensitivities strongly predict periods of financial crisis.  Moreover, the pattern of temporal changes in correlation structure between volatility and returns is readily interpretable in terms of the major events of the periods in question.

\vspace{0.1in}

\noindent Keywords: contagion, financial crises, sequential Monte Carlo, stochastic volatility
\end{abstract}

\newpage

\doublespacing

\section{Contagion and financial crisis}

Contagion refers to the idea that asset returns in inter-related financial markets sometimes exhibit anomalous patterns of correlation. Yet these anomalies are difficult even to identify, much less characterize empirically, and their precise causes---whether financial, macroeconomic, or behavioral---are hotly debated.

In this paper, we study the period of presumptive financial contagion associated with the European sovereign-debt crisis of 2008--2011, which arose out of market fears that Greece would default on its sovereign debt.  The crisis soon grew to include fears about the future solvency of other European nations with large debt loads.  These fears, and the associated turmoil in global financial markets, led to the creation of the European Financial Stability Fund (EFSF) in May of 2010, which was intended to faciliate low-cost loans for other struggling EU members, including Portugal and Ireland.  Yet uncertainty continued to cloud the global macroeconomic outlook throughout 2011 and early 2012, a period during which the likely impacts of Eurozone contagion on financial markets, monetary and fiscal policy, and trade links were far from known.

During this period of instability, stock returns in both European and US markets exhibited explosive daily movements that are very hard to explain using traditional asset-pricing models.  To understand the magnitudes of the effects that have to be modeled, consider that the German DAX index fell from a high of 7402 at the beginning of July 2011 to 5216 by 4 October, resulting in a -30\% return.  Over the same period the UK FTSE100 had a fall of similar magnitude.  From a volatility perspective, the S\&P 500 volatility index (or VIX) dramatically increased from 15\% on July 1, 2011 to 45.5\% by 4 October, staying persistently high for at least a week thereafter.  

The sheer length of the European debt crisis provides a natural testing ground for models of contagion.  We take an asset-pricing perspective on this problem, and propose a multifactor, explosive-stochastic-volatility modeling framework.  This approach generalizes existing models by allowing for the possibility of explosive, mutually exciting shocks to volatility that play a role in the cross-section of expected returns.  Our work therefore builds on the long line of research into stochastic volatility and inference for Bayesian state-space models more generally, including \citet{gamerman:migon:1993} \citet{kim:shep:chib:1998}, \citet{AguilarWest00}, \citet{liu:west:2001}, \citet{jaqu:pols:ross:2002}, and \cite{Chib04}.  Other important references on Bayesian computation for non-linear state-space models include \citet{carvalho:etal:2010}, \cite{niemi:west:2010}, and \citet{lopes:tsay:2011}, among others.

The closest related model is that of \citet{lopes:migon:2002}, who also incorporate stochastic volatility into an asset-pricing model for contagion.  The crucial difference in our model is that it can express all three ways in which market shocks tend to cluster during times of crisis: \textit{time-series} clustering, where large shocks today correlate with large shocks tomorrow; \textit{cross-sectional} clustering, where large shocks in one region correlate with large shocks in other regions; and \textit{directional} clustering, where shocks to aggregate market volatility correlate with specific directional biases in contemporaneous country-level returns.

These first two forms are, respectively, the ``heat wave'' and ``meteor shower'' metaphors of volatility (or jump) clustering proffered by \citet{engle:ito:li:1990}.  \citep[See also][]{eraker:johannes:polson:2002,aitsahalia:etal:2010}  These forms of clustering are certainly not unique to contagion periods.  Our approach differs from this line of thought, in that we emphasize the role of directional clustering, driven by aggregate volatility shocks, in explaining contagion episodes.  To cite one example: during the European sovereign debt crisis of 2010, our analysis finds that aggregate volatility shocks were systematically associated with large positive residual returns on the Spanish stock market, and large negative residual returns on the German market.  Existing volatility models are agnostic with respect to the sign of these residuals, and provide no mechanism for aggregate volatility shocks to be associated with specific directional biases in country-level moves.  Yet such biases show up clearly in the data.

\begin{figure}
\begin{center}
\includegraphics[width=5.5in]{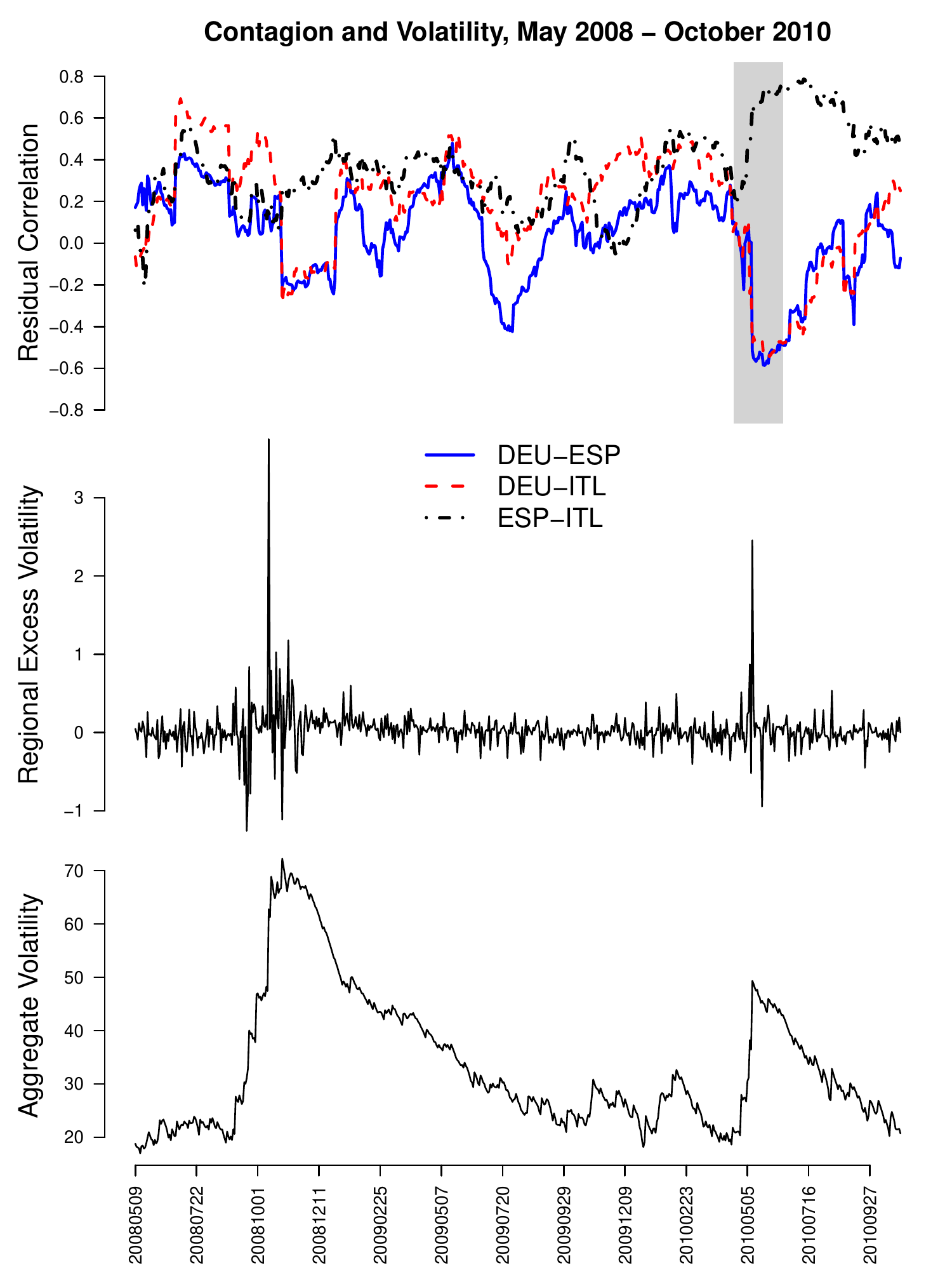}
\end{center}
\caption{\label{fig:pairwise-correlation} Top: trailing two-month correlation coefficients for the residuals of Germany--Spain, Germany--Italy, and Italy--Spain from a two-factor model.  Middle: our proposed risk factor that measures regional excess volatility.  Bottom: estimate of global market volatility that arises from our explosive stochastic-volatility model.}
\end{figure}

We analyze daily data on returns from country-level equity indices during the European sovereign debt crisis, and reach three main empirical findings.  First, we present evidence that European equity markets exhibited significant excess correlation during the debt crisis of 2010, relative to an asset-pricing model that assumes regional market integration.  Figure \ref{fig:pairwise-correlation} gives a brief visual summary of this evidence.  It depicts the trailing two-month residual correlation coefficients among Germany, Italy, and Spain between May 2008 and October 2010.  The residuals are from a two-factor model incorporating global and European market-risk factors, described in Section \ref{sec:contagion-empirical}.  In April and May of 2010 (shaded grey in the figure), there is a clear divergence from the historical norm, precisely coinciding with the Greek sovereign-debt crisis and associated bailout.

Second, we find that part of this excess correlation can be attributed to the impact of mutually exciting volatility shocks in the cross-section of expected returns.  We propose global and regional volatility-risk factors to quantify this impact, and show how to construct these volatility risk factors directly from the time series of market returns.  This is similar to the work of \citet{ang:hodrick:JOF:2006}, but differs in two key respects: we consider returns at the level of national portfolios, rather than individual domestic equities; and we construct the volatility risk factors directly from the time series of market returns, rather than indirectly from option prices.

Our results show that markets have time-varying sensitivities to these risk factors, which implies time-varying correlation between shocks to volatility and contemporaneous shocks to returns.  Crucially, periods of increased sensitivity to these risk factors coincide with periods of excess correlation.  It is only a slight oversimplification of our findings to say that volatility shocks provide little predictive value during quiet periods, but significant predictive value during times of market crisis, which are precisely the times when market shocks exhibit strong directional clustering.  Figure \ref{fig:pairwise-correlation} shows this correspondence: the regional volatility-risk factor accurately predicts periods during which correlations among the residuals for Germany, Italy, and Spain diverge significantly from their long-run average.  The pattern of volatility factor loadings that we find during crisis periods, moreover, is readily interpretable.  For example, during May 2010, the regional volatility factor partially accounts for two notable phenomena: Spanish, Italian, and Belgian markets rallying in anticipation of a Greek bailout; and German and British markets suffering in anticipation of footing the bill.

Our third main finding concerns the contemporaneous relationship between volatility and expected returns.  This finding is tangential to our main point, but is of independent interest in the finance literature.  We are primarily interested in the role of volatility shocks in understanding contagion, rather than in estimating volatility \textit{per se}.  Nonetheless, in performing a formal statistical assessment of our volatility model, we find a significant negative relationship between daily volatility and expected returns on the U.S.~market portfolio.  Crucially, no such finding on the daily time scale can be supported using other, more traditional volatility models.  This finding links our methodology with the longstanding debate on the relationship between risk and return \citep[e.g.][]{glosten:jag:runkle:1993}.

Our empirical approach relies upon a new method for measuring volatility shocks, which is a necessary step in constructing the volatility risk factors that we propose.  As we will show, traditional volatility estimators cannot explain the large shocks to volatility seen during recent financial crises, and sometimes lag several days behind during times of acute turmoil.  Many such large volatility spikes are evident in Figure \ref{fig:pairwise-correlation}.  This leaves investors in the unsatisfying position of accepting that asset prices undergo moves of 6, 8, or even 10 standard deviations far more often than their models would predict.  To correct this problem, we propose an explosive stochastic volatility (ESV) model, where the random shocks to aggregate volatility arise from a distribution whose tails are much heavier than Gaussian, and which can be fit straightforwardly using particle learning.  This model does not require implausibly large shocks in order to generate observed movements in asset returns.

The rest of the paper proceeds as follows.  In Section 2, we describe the role of volatility in understanding presumptive periods of financial contagion.  In this context we review several, related proposals that motivate our decision to focus on how volatility shocks enter the cross-section of expected returns, providing important context for our proposed asset-pricing models.  In Section 3, we explain the most general version of our mutually exciting stochastic volatility model, wherein the correlation between volatility shocks and return shocks are explicitly quantified.  We also describe a handful of empirical simplifications that we make in order to fit the model to country-level European returns.  Our estimation strategy then unfolds in two stages.  In Section 4, we estimate both the expected returns and the daily volatility states for both the U.S. and European markets using particle learning.  We also describe the results of an extensive model assessment, wherein we benchmark our volatility model against an alternative model based on the GARCH framework.  In Section 5, we use the filtered mean and volatility states from Section 4 to fit the encompassing model described in Section 3.  Sectino 6 concludes with some final remarks about how our findings fit into the existing literature.

\section{The role of volatility in understanding contagion}

Existing economic theory provides several reasons why the study of market volatility---the primary modeling focus of our paper---would play a major role in understanding contagion.  We divide these into two broad categories: economic (or fundamentals-based) reasons; and statistical reasons.

Many previous authors have provided economic interpretations of why volatility shocks would predict variation in asset returns during crisis periods, even when other risk factors have been priced.  One explanation is that volatility shocks are simply a proxy for factors that can be loosely grouped into the category of investor behavior \citep[e.g.][]{dornbusch:2000}.  A second possible interpretation can be found in the literature on the portfolio flows of large international investors.  \citet{froot:etal:2001} study the relationship between local market returns and capital inflows from large institutional investors domiciled outside the local market.  They find evidence that these inflows are typically associated with larger expected returns, but that transitory inflows predict lower returns.  To the extent that large regional volatility shocks predict increasingly transitory capital flows, they may also predict returns.  

A third possible interpretation is that volatility shocks are a proxy for changes in the informational efficiency of equity markets.  The resulting effects may take several forms.  \citet{calvo:mendoza:2000} observe that, due to the potentially high cost of procuring and assimilating country-specific information, investors may rationally engage in herding behavior due to mutual dependence upon a relatively small cohort of information conduits.  Additionally, standard models of market microstructure suggest that investors with informational advantages will accrue their desired market positions slowly, in order to minimize transaction costs.  Therefore, the portfolio flows of informed investors will be more highly autocorrelated than those flows corresponding to uninformed transactions, and therefore more strongly predictive of future returns.  If informational asymmetries are high, flows involving investors with an informational advantage will constitute a larger fraction of total flows.  Under this assumption, if volatility shocks correspond both to information asymmetries and to increased trading volume, they will also enter the cross-section of returns.

Moreover, stochastic volatility offers perhaps the best modeling framework for understanding contagion when accurate data on macroeconomic fundamentals is lacking.  The absence of data on fundamentals is especially acute when studying returns on a daily scale, rather than a scale of months or weeks.  Working with daily data poses unique challenges: the potentially large size of daily volatility moves, the poor signal-to-noise ratio, and the fact that different markets operate on different trading calendars.  The models we use in this paper directly address the first two issues.  To address the third, we use daily data on U.S.-based exchange-traded funds (ETFs) corresponding to major European equity indices, rather than the indices themselves.  The use of market proxies constructed from ETFs (which are highly liquid and free of any apparent arbitrage on a daily time scale) eliminates the complications due to differences in the trading calendar across different nations.

Using data on coarser time scales potentially allows important information such as trade variables and interest rates to be incorporated into the asset-pricing models, as in \citet{bekaert:harvey:ng:2005}.  This allows for a more precise economic interpretation of any contagion effects that might be found.  Yet low-frequency data imposes an even larger disadvantage, given the poor statistical precision it allows in separating contemporaneous from time-lagged covariation.  Since trade conditions and other macroeconomic variables do not change fast enough to drive much of the variation in daily returns, we are left with volatility shocks as the most salient estimable quantity to use in our pricing model.

Finally, there are purely statistical issues associated with identifying contagion.  The most difficult of these are related to the issue of time-varying volatility.  Suppose that one has specified an asset-pricing model that captures all relevant shared risk factors for a group of assets, such that the idiosyncratic components of variation are uncorrelated for all assets.  Even under this ideal scenario, a deceptive illusion of excess correlation may arise if the presence of time-varying volatility is ignored.

Such an effect can be seen in the context of an international version of the capital asset pricing model (CAPM) with time-varying volatility.  Suppose that $y_{it}$ is the excess return of the national equity index for nation $i$ at time $t$; that $\mu_{t}$ is the conditional expected excess return of the world market, given information available at time $t-1$; that $x_t$ is the realized excess return on the world market; and that $x_t = \mu_t + e_t$, where $\var(e_t) = \sigma^2_t$.  If markets are globally integrated in accord with the world CAPM, then
\begin{equation}
\E(y_{it}) = \beta_i \mu_{t} \, ,
\end{equation}
where the residuals $\epsilon_{it} = (y_{it}-\beta_i \mu_{t})$ have variance $\sigma^2_{it}$.

Even in this simple setting, ignoring the effect of time-varying volatility can lead to spurious findings of excess correlation.  This happens because cross-market correlations are conditional upon the overall market volatility $\sigma_t^2$, since the market shock $\epsilon_t$ affects the return for every asset.  Hence when market volatility is high, estimates of these correlations are biased upward \citep[see, e.g.][]{forbes:rigobon:2002}.  A number of authors have attempted to correct for this in studying previous crises previous crises.  For example, \citet{bekaert:harvey:ng:2005} explore the importance of segmentation versus integration in explaining contagion, and consider the possibility that increased volatility of regional risk factors can account for excess correlation during the 1997 Asian crisis and the 1994 Mexican crisis.

The problem is further exacerbated by an errors-in-variables effect.  Suppose that one were to fit the CAPM using the realized market returns $x_t$.  Then the least-squares residuals corresponding to assets $i$ and $j$ will be correlated, even if all variance terms are known, and even if the real $\epsilon_{it}$'s are cross-sectionally independent.  This excess correlation, moreover, will be higher for those residuals corresponding to times of market stress, when $\sigma^2_t$ is large.  To see this informally, note that
\begin{eqnarray*}
\mbox{cor}(y_{it}, y_{jt}) &=& \mbox{cor}\{\beta_i(x_t - e_t) + \epsilon_{it}, \beta_j (x_t - e_t) + \epsilon_{jt}\} \\
&=& \frac{\beta_i \beta_j }{ \big( \beta_i^2+ \sigma_{it}^2 / \sigma_t^2 \big)^{1/2} \cdot \big( \beta_j^2  + \sigma_{jt}^2 / \sigma_t^2  \big) ^{1/2} } \, ,
\end{eqnarray*}
which is large when $\sigma^2_t$ is large relative to the idiosyncratic variances $\sigma^2_{it}$.  This excess residual correlation results from two well-known facts: that the least-squares estimators for $\beta_i$ and $\beta_j$ are biased downwards when the expected market return is observed with error; and that this bias is more severe when $\sigma^2_t$ is large.

Therefore, even asset-price movements that arise from the most basic form of the CAPM can look suspiciously like excess correlation, unless both the errors-in-variables problem and the time-varying volatility problem are handled appropriately.  This kind of correlation is best thought of as correlation arising from integrated markets, and thus is not really contagion according to the asset-pricing definition used in this paper.

\section{Mutually exciting stochastic volatility}
\label{sec:volatility} 

\subsection{Motivation}

We therefore adopt the goal of quantifying the effect of volatility on market returns during presumptive contagion periods, remaining agnostic on the causal interpretation of these effects.  We focus most specifically on the phenomenon of ``directional clustering'' referred to in the introduction: namely, that a large shock to aggregate volatility predicts specific biases in the signs of country-level returns. Figure \ref{fig:pairwise-correlation} hints at precisely this effect, when, for example, the May 2010 spike in Eurozone volatility was associated with large, historically aberrant negative correlations between Germany and Spain.

Existing models do not explicitly address this effect, whereby epochs of high volatility are associated with specific directional biases in country-level asset returns.  For example, a traditional multivariate stochastic-volatility model asserts that a large positive residual for the Italian stock market during May of 2010 was equally likely to be associated with a large positive residual for Germany as it was a large negative residual.  Figure \ref{fig:pairwise-correlation} puts the lie to this assumption.  In contrast, our goal is to characterize these directional biases explicitly, and to understand the role they play in presumptive contagion episodes.  

\subsection{Modeling framework}

Let $y_{it}$ represent the excess returns for asset $i$ during period $t$.  A general, dynamic multifactor representation of expected returns would take the following form:
\begin{eqnarray}
y_{it} &=& \alpha_{it} + \sum_{j=1}^K \beta^{(j)}_{it} (f_{jt} - \mu_{jt}) + \epsilon_{it} \, .
\end{eqnarray}
Here $f_{jt}$ represents a shared risk factor, with $\mu_{jt}$ denoting the corresponding conditional mean of that factor.  For example, the CAPM sets $K=1$, $f_{t}$ as the market return, and $\mu_t$ as the expected market return.

Our model specification accounts for the conditional volatility of the risk factors themselves, taking the square-root volatility model of \citet{heston:1993} as a starting point.  Throughout the following exposition, we assume that all volatility states are restricted so that the square-root volatility process never becomes negative.

We describe the most general formulation of our model, before considering specific simplifications that are necessary in order to fit the model to data.  In matrix form, we assume that
\begin{eqnarray}
\beff_t &=& \bmu_t + U_t \bxi_t  \\
\by_t &=& \balpha_t + B_t (\beff_t - \bmu_t) + V_t \beps_t \label{eqn:observation1} \, ,
\end{eqnarray}
where $\beff_t = (f_{1t}, \ldots, f_{Kt})'$ is a vector of shared risk factors at time $t$; $B_t$ is a matrix of factor loadings; $\by_t = (y_{1t}, \ldots, y_{pt})'$ is a $p$-vector of excess returns for individual national equity markets; and where $U_t = \mbox{diag}(u_{1t}, \ldots, u_{Kt})$ and $V_t = \mbox{diag}(v_{1t}, \ldots, v_{pt})$ are diagonal matrices of volatility states (i.e.~conditional standard deviations).

We assume a multivariate stochastic volatility model for the complete vector of volatility states $(\bu_t, \bv_t) = (u_{1t}, \ldots, u_{Kt}, v_{1t}, \ldots, v_{pt})'$:
\begin{equation}
\label{eqn:volatilitystates}
\left(
\begin{array}{c}
\bu_t \\
\bv_t 
\end{array}
\right) 
=
\left(
\begin{array}{c}
\bu_0 \\
\bv_0 
\end{array}
\right) 
 + 
  \left(
\begin{array}{cc}
\Phi_{uu} & 0 \\
0 & \Phi_{vv}
\end{array}
\right) 
 \left(
\begin{array}{c}
\bu_{t-1} \\
\bv_{t-1} 
\end{array}
\right) 
 +  \left(
\begin{array}{c}
\boldeta^u_{t} \\
\boldeta^v_{t} 
\end{array}
\right) \, ,
\end{equation}
where $\boldeta_t = (\boldeta^u_t, \boldeta^v_t)'  $ is a vector of longitudinally independent innovations to volatility.  See, for example, \citet{eraker:johannes:polson:2002} and \citet{Chib04}. 

Finally, define $\bpsi_t = ( \bxi_t,  \beps_t, \boldeta_t)'$ to be the stacked cross-sectional vector of innovations at time $t$ in the factors, returns, and volatility states, respectively.  Our model is completed by specifying a covariance structure for these innovations, which we assume to be multivariate Gaussian:
$$
\bpsi_t \sim \N(0, \Sigma_t) \, .
$$
Suppressing the time index for ease of notation, partition $\Sigma$ as
$$
\Sigma = \left(
\begin{array}{cccc}
\Sigma_{\xi \xi} 			& \Sigma_{\xi \epsilon} 		& \Sigma_{\xi u} 		& \Sigma_{\xi v} \\
\Sigma_{\xi \epsilon} ' 	& \Sigma_{\epsilon \epsilon} 	&  \Sigma_{\epsilon u}	& \Sigma_{\epsilon v} \\
 \Sigma_{\xi u}'	 		& \Sigma_{\epsilon u}'		& \Sigma_{u u} 		& \Sigma_{u v}  \\
 \Sigma_{\xi v}'			& \Sigma_{\epsilon v}' 		& \Sigma_{u v}'  		& \Sigma_{v v} 
\end{array}
\right) \, .
$$

Some further identifying restrictions on $\Sigma_t$ are necessary, due to the presence of the diagonal scale factors $U_t$ and $V_t$ in the state evolution equations, and because the factor innovations also explicitly contribute to the observation equation (\ref{eqn:observation1}), since $\beff_t - \bmu_t = U_t \bxi_t$.  To address these, we assume that:
\begin{enumerate}
\item $\Sigma_{\xi \xi}$ and $\Sigma_{\epsilon \epsilon}$ are correlation matrices; and
\item the factor innovations $\bxi_t$ are cross-sectionally independent of the return innovations $\beps_t$ and the volatility innovations $\boldeta_t$, so that $\Sigma_{\xi \epsilon} = \Sigma_{\xi u} = \Sigma_{\xi v} = 0$.
\end{enumerate}

We are therefore left with a simplified model where
$$
\Sigma = \left(
\begin{array}{cccc}
\Sigma_{\xi \xi} 			& 0 		& 0 		& 0 \\
0	& \Sigma_{\epsilon \epsilon} 	&  \Sigma_{\epsilon u}	& \Sigma_{\epsilon v} \\
0		& \Sigma_{\epsilon u}'		& \Sigma_{u u} 		& \Sigma_{u v}  \\
0		& \Sigma_{\epsilon v}' 		& \Sigma_{u v}'  		& \Sigma_{v v} 
\end{array}
\right) \, ,
$$
with inverse partitioned conformally as
$$
\Sigma^{-1} = \Omega =  \left(
\begin{array}{cccc}
\Omega_{\xi \xi} 			& 0 		& 0		& 0 \\
0 	& \Omega_{\epsilon \epsilon} 	&  \Omega_{\epsilon u}	& \Omega_{\epsilon v} \\
0	 		& \Omega_{\epsilon u}'		& \Omega_{u u} 		& \Omega_{u v}  \\
0			& \Omega_{\epsilon v}' 		& \Omega_{u v}'  		& \Omega_{v v} 
\end{array}
\right) \, .
$$
Even with these identifying restrictions, the resulting model is still very flexible.  Importantly, it can accommodate all three forms of clustering described above.
\begin{description}
\item[Cross-sectional:] A large shock in one market can propagate to other markets in two ways.  First, volatility shocks are themselves cross-sectionally correlated at level of factors ($\Sigma_{uu}$), returns ($\Sigma_{vv}$), and factor-return interactions ($\Sigma_{uv}$).  Second, time-lagged cross-sectional correlations can be amplified by the regression matrix $D$ in the volatility-state evolution equation.
\item[Time-series:] Equation (\ref{eqn:volatilitystates}) descibes a state-evolution equation wherein volatility states are auto- and cross-correlated over time.
\item[Directional:] Unless both $ \Sigma_{\epsilon u}$ and $ \Sigma_{\epsilon v}$ are identically zero, shocks to volatility will be associated with specific biases in the signs of the return residuals ($\beps_t$).
\end{description}
The theory of the previous section predicts that periods of contagion would be associated with changes in $ \Sigma_{\epsilon u}$ and $ \Sigma_{\epsilon v}$---that is, the correlation between shocks to volatility and shocks to returns.  We are now in a position to formally investigate this prediction.

\subsection{Estimation strategy}

We view the general dynamic model we have just outlined as a useful conceptual framework for thinking about contagion.  Nonetheless, it poses a number of difficult statistical issues associated with high-dimensional state-space models \citep[see, e.g.][]{liu:west:2001,carvalho:etal:2010}.  These include the unknown parameters in the state evolution equations; the large dimension of the full covariance matrix $\Sigma_t$; the potentially heavy-tailed character of the volatility shocks; the errors-in-variables problem that arises because the conditional factor means $\bmu_t$ are unobserved; and the need to specify a model for how parameters such as $B_t$ change over time. 

Therefore, rather than attempting to solve all of these issues in the context of a single dynamic linear model, we make a number of specific simplifications for the purpose of fitting this model to data from the European sovereign debt crisis.  These simplifications, and our two-stage estimation strategy, are guided by the desire to test the main prediction of the full model: namely, that innovations in the volatility sequence enter the cross-section of returns ($\by_t$), above and beyond their contribution to the volatility of the factors ($\beff_t)$.

To see this, suppose we are given the values of the volatility innovations $\boldeta_t$.
Appealing to standard multivariate-normal theory, the conditional distribution of the return innovations $\beps_t$ is
$$
(\beps_t \mid \boldeta_t) \sim \N \left( m_t^{\epsilon}, H^{\epsilon}_t \right) \, ,
$$
where
\begin{eqnarray}
H^{\epsilon}_t &=& \Sigma_{\epsilon \epsilon} - ( \Sigma_{\epsilon u} \  \Sigma_{\epsilon v})
\left(
\begin{array}{cc}
\Sigma_{u u} 		& \Sigma_{u v}  \\
 \Sigma_{u v}'  		& \Sigma_{v v} 
\end{array}
\right)^{-1}
 ( \Sigma_{\epsilon u} \  \Sigma_{\epsilon v})'  \\
m_t^{\epsilon} &=& \Omega_{\epsilon \epsilon}^{-1} \Big ( \Omega_{\epsilon u} \  \Omega_{\epsilon v} \Big)
\left(
\begin{array}{c}
\boldeta^u_{t} \\
\boldeta^v_{t} 
\end{array}
\right)
\end{eqnarray}
%Appealing to standard multivariate-normal theory, the conditional expectation $m_t^{\epsilon}$ of the return innovations $\beps_t$ is
%\begin{eqnarray}
%m_t^{\epsilon} &=& \Omega_{\epsilon \epsilon}^{-1} \Big ( \Omega_{\epsilon u} \  \Omega_{\epsilon v} \Big)
%\left(
%\begin{array}{c}
%\boldeta^u_{t} \\
%\boldeta^v_{t} 
%\end{array}
%\right)
%\end{eqnarray}

Therefore we may write the conditional model for returns as
\begin{equation}
\label{eqn:conditionalmodel1}
\by_t = \balpha_t + B_t (\beff_t - \bmu_t) + \Gamma^u_t \boldeta^u_t +  \Gamma^v_t \boldeta^v_t  + V_t \be_t \, ,
\end{equation}
where
\begin{eqnarray*}
\Gamma^u_t &=& V_t \Omega^{-1}_{\epsilon \epsilon, t} \Omega_{\epsilon u, t} \\
\Gamma^v_t &=& V_t \Omega^{-1}_{\epsilon \epsilon, t} \Omega_{\epsilon v, t} 
\end{eqnarray*}
are matrices of regression coefficients, and where $\var(\be_t) = H^{\epsilon}_t$.

Observe that, in the conditional model of Equation (\ref{eqn:conditionalmodel1}), the volatility innovations behave like shared risk factors that enter the cross section of returns.  We can therefore describe the model in either of two equivalent ways: 1) time-varying correlation between residuals in the returns and volatility evolution equations, described by blocks $\Omega_{\epsilon u, t}$ and $\Omega_{\epsilon v, t}$ of the overall precision matrix $\Omega_t = \Sigma_t^{-1}$; or 2) time-varying sensitivities to volatility-risk factors, described by loadings $\Gamma^u_t$ and $\Gamma^v_t$.

We work directly with this second (conditional) model, subject to several further simplifications.  First, we assume a fixed-parameter, two-factor model that describes daily excess returns in terms of exposure to global-market and regional-market risk factors, along the lines of \citet{bekaert:harvey:ng:2005}.  This leads to the following marginal model for the returns series:
$$
y_{it} = \alpha_i + \beta_i^{US} (f^{US}_{t} - \mu^{US}_{t}) + \beta_i^{EU} (f^{EU}_{t} - \mu_{t}^{EU}) + \epsilon_{it} \, ,
$$
where $y_{it}$ is the excess return for index $i$ on day $t$; $f^{US}_t$ ($\mu^{US}_t$) is the observed return (expected return) on the value-weighted US market portfolio; and $f^{EU}_t$  ($\mu^{EU}_t$) is the observed return (expected return) on the pan-European index from Morgan Stanley Capital International, which corresponds to a value-weighted portfolio drawn from the 16 largest equity markets in Europe.  We use the US market return as a proxy for the global market.  We also use the return Vanguard's European index exchange-traded fund (ticker VGK) as a proxy for the MSCI European index.  Since this ETF is traded in the US market, it allows us to sidestep differences in trading calendars among the individual European markets themselves.  

Second, we assume that the full covariance matrix $\Sigma$ (which describes all correlation among the shocks to factors, returns, and volatility states) is restricted in such a way that $\Omega_{\epsilon v} = 0$ for all $t$.  This imposes the requirement that shocks to country-level returns are conditionally independent of shocks to country-level volatility, given the shocks to factor volatility.  We have therefore specified a particular form of Gaussian graphical model for $\Sigma$ \citep[see, e.g.][]{carvalhowest2007}.  Accordingly, $\boldeta^v_t$ drops out of the conditional model (\ref{eqn:conditionalmodel1}) for returns, with only $\boldeta^u_t$ remaining (in addition to the factors themselves).

Third, we assume that the innovations to country-level volatility and factor-level volatility are conditionally independent, given all other contemporaneous parameters and states of the DLM.  This encodes the assumption that the factor-level volatility innovations are sufficient to describe the correlation between volatility and returns, and that the country-level volatility shocks offer little extra information.

Fourth, we assume that the factor volatilities $(u_t^1, u_2^t)$ for the US and EU markets, respectively, obey the following triangular system:
\begin{equation}
\label{eqn:volatilitytriangle}
\left(
\begin{array}{c}
u_t^{1} \\
u_t^{2} 
\end{array}
\right) 
=
\left(
\begin{array}{c}
u^{1}_0 \\
u^{2}_0 
\end{array}
\right) 
 + 
  \left(
\begin{array}{cc}
\phi_1 & 0 \\
0 & \phi_2
\end{array}
\right) 
\left(
\begin{array}{c}
u_{t-1}^{1} \\
u_{t-1}^{2} 
\end{array}
\right) 
 +   \left(
\begin{array}{cc}
l_{11} & 0 \\
l_{21} & l_{22}
\end{array}
\right) 
 \left(
\begin{array}{c}
\delta^1_{t} \\
\delta^2_{t} 
\end{array}
\right) \, ,
\end{equation}
where $\delta_t^1$ and $\delta_t^2$ are mutually independent.  We interpret $\delta_1$ as a shock to global market volatility, and $\delta_t^2$ as an independent shock to region-level volatility.  The actual volatility in the regional market-risk factor is then a linear combination of $\delta_t^1$ and $\delta^2_t$, since from (\ref{eqn:volatilitystates}),
$$
\boldeta_t^v = \left(
\begin{array}{cc}
l_{11} & 0 \\
l_{21} & l_{22}
\end{array}
\right) 
 \left(
\begin{array}{c}
\delta^1_{t} \\
\delta^2_{t} 
\end{array}
\right) \, .
$$
This implies that the lower-triangular matrix $L$ is the Cholesky factor for the marginal covariance matrix of the factor innovations $\boldeta_t^v$.

Finally, we assume that shocks to returns are conditionally independent of each other, given the factors, volatilities, and volatility innovations.  Statistically speaking, this means that $\Omega_{\epsilon \epsilon}$ is assumed to be a diagonal matrix.

We are therefore left with a simplified form of the full model wherein
$$
\Sigma = \left(
\begin{array}{cccc}
\Sigma_{\xi \xi} 			& 0 		& 0 		& 0 \\
0	& \Sigma_{\epsilon \epsilon} 	&  \Sigma_{\epsilon u}	&0 \\
0		& \Sigma_{\epsilon u}'		& \Sigma_{u u} 		& 0 \\
0		& 0		& 0 		& \Sigma_{v v} 
\end{array}
\right) 	\quad 	\mbox{and}	 \quad  \Omega =  \left(
\begin{array}{cccc}
\Omega_{\xi \xi} 			& 0 		& 0		& 0 \\
0 	& \Omega_{\epsilon \epsilon} 	&  \Omega_{\epsilon u}	& 0 \\
0	 		& \Omega_{\epsilon u}'		& \Omega_{u u} 		& 0 \\
0			& 	0	& 0  		& \Omega_{v v} 
\end{array}
\right) \, .
$$
This, in turn, implies that the conditional model for returns (\ref{eqn:conditionalmodel1}) reduces to
\begin{equation}
\label{eqn:conditionalmodel2}
\by_t = \balpha + B (\beff_t - \bmu_t) + \Gamma^u  \boldeta^u_t + V \be_t \, ,
\end{equation}
where the $\be_t$'s are independent of $\boldeta_t^u$ with variance
$$
\var(\be_t) = \Omega_{\epsilon \epsilon}^{-1} =  \Sigma_{\epsilon \epsilon} -  \Sigma_{\epsilon u} \Sigma_{u u}^{-1}  \Sigma_{\epsilon u}' \, .
 $$

These simplifications suggest that, if one knew the volatility states, it would be possible to estimate the time-varying pattern of correlation between volatility shocks and return shocks, strictly by estimating $\Gamma^u$.  This leads us to adopt the following two-stage approach for estimating the model.
\begin{enumerate}
\item Fit independent heavy-tailed stochastic volatility models to the US and EU markets, respectively, for the purpose of getting filtered estimates for the conditional means and volatilities of the factors.  This is described in the next section.  
We use these filtered states to regress the EU volatility innovations on the US volatility innovations in order to extract the two independent sets of volatility risk factors.  
\item Then estimate the conditional model in (\ref{eqn:conditionalmodel2}), fitting a traditional factor model using the volatility innovations as shared risk factors.  By using $\delta^1_t$ and $\delta_t^2$ rather than $\eta_t^1$ and $\eta_t^2$, we eliminate collinearity among the predictors.
\end{enumerate}

\section{Estimating time-varying market volatility}

\subsection{The relationship between risk and return}

Our first, intermediate goal is to model the conditional mean and volatility of the daily return on the US and EU markets.  This allows us to isolate the global and regional market shocks, and to extract a volatility risk factor for incorporating into the pricing model for the market returns of individual national equity indices (\ref{eqn:conditionalmodel2}).  It is also essential if we are to correct for the errors-in-variables effect that arises because the factor means $\bmu_t$ in this model are unknown.

Extracting a volatility-risk factor requires specifying the the fundamental relationship between risk and expected return on the market. The market models we consider are simple and fairly standard, and are identical to those used in the literature on the equity risk premium.  Yet as we will see, they are still tricky to estimate.

Following \citet{merton:1980}, \cite{glosten:jag:runkle:1993}, and others, we consider three baseline models for the excess market return.  The first two of these models also have special ``no-intercept'' cases, which we denote Models 1(b) and 2(b), respectively, giving five overall models.  To avoid confusion with the notation already established, we phrase the issue somewhat generically, letting $x_t$ denote the market return and $\sigma_t$ the conditional standard deviation of the market return.  The models we consider are:
\begin{align*}
\text{Model 1(a):} \; \; x_t & = \alpha_0 + \alpha_1 \sigma^2_t + \epsilon_t \ , \ \var(\epsilon_t) = \sigma^2_t \\
\text{Model 2(a):} \; \; x_t & = \alpha_0 + \alpha_1 \sigma_t + \epsilon_t \ , \ \var(\epsilon_t) = \sigma^2_t \\
\text{Model 3:} \; \; x_t & = \alpha_0 + \epsilon_t \ , \ \var(\epsilon_t) = \sigma^2_t  
\end{align*}
Model 3 hypothesizes that expected market returns are constant over time, and are unrelated to the conditional volatility of the market.  We also entertain Model 1(b), the special case of 1(a) where $\alpha_0 = 0$.  This approximates an equilibrium model in which the relative risk-aversion function for the representative investor is constant over long periods of time.  Likewise, we entertain Model 2(b), the special case of 2(a) where $\alpha_0 = 0$.  This restriction implies that the market price of risk, $\E(x_t)/\sigma_t$, is constant over long periods of time.  

Rational, risk-averse investors require higher expected returns to compensate them for taking on greater levels of risk.  Yet this need not imply a positive relationship between risk and return over long time periods.  For example, if investors systematically tend to save more of their wealth during risky times, then asset prices may be bid up considerably, and expected excess returns may go down as a result.  For this and other reasons, the sign of $\alpha_1$ may be either positive or negative without contradicting the assumption that investors are rational.  See \cite{glosten:jag:runkle:1993} for an overview of the literature on the relationship between market volatility and the equity risk premium.

These models have a relatively long history in finance, where the need to account for time-varying volatility in characterizing the expected market return has long been recognized.  For example, \citet{merton:1980} considers Models 1(b), 2(b), and 3 for monthly data, using very simple estimates of monthly volatility.  Merton concludes his study by observing that ``[the] most important direction is to develop accurate variance estimation models which take account of the errors in variance estimates.''  \cite{glosten:jag:runkle:1993} do precisely this.  Focusing on Model 1(a), they use an asymmetric GARCH model incorporating the risk-free interest rate and seasonal dummies to estimate monthly volatility.  They find evidence that $\alpha_1$ is significantly negative, implying a negative conditional correlation between returns and volatility.

Many further elaborations of this basic framework are possible, but typically necessitate working on a coarser time scale.  \citet{bekaert:harvey:ng:2005}, for example, incorporate many economic fundamentals into their model of the the expected market return.  We do not incorporate these fundamentals into our models because they change very slowly on a daily time scale and will therefore drive very little of the variation.

Each of the models above requires further assumptions about the behavior of the volatility term $\sigma_t$.  We assume a square-root stochastic volatility model with heavy-tailed shocks:
\begin{eqnarray}
\sigma_t &=& \sigma_0 + \phi \sigma_{t-1} + \lambda_t z_t \\
\lambda_t &\sim& t^{+}_{\nu} \\
z_t &\sim& \N(0,\tau^2) \, ,
\end{eqnarray}
with $\sigma_t$ truncated below at zero, and where $t^{+}_{\nu}$ is a positive-$t$ random scale with $\nu$ degrees of freedom.

This differs from a traditional Gaussian SV model, in that the innovation is itself a product of two terms: a normal shock (which is present in a normal SV model), and a half-t-distributed ``local'' scale factor (which isn't).  This results in a horseshoe distribution (when $\nu=1$) for the marginal of the shock $\lambda_t z_t$ \citep{Carvalho:Polson:Scott:2008a}, with tails far heavier than Gaussian.  Formally, it corresponds to a discretized version of subordinated Brownian motion for the square-root volatility process, a construction first explored by S.~Bochner in the 1950's.  The increments of the random clock, or subordinator, are half-$t$ distributed.

The intuition behind our approach is that the daily volatility of asset returns is highly persistent during normal times, but on rare occasions will change drastically and rapidly.   When these drastic moves occur, rational investors must quickly discount the past and revise their expectations about volatility. 

Fat-tailed versions of traditional stochastic volatility models have been proposed to understand volatility filtering \cite{jaqu:pols:ross:2002}
and currency returns in a multi-factor setting \cite{AguilarWest00}. Our approach here is different in that we explore the expected-return
relationship with time-varying volatility. This allows us to provide a volatility factor that also respects the fundamental risk-return trade-off 
inherent in an asset-pricing market model.  We also fit the model with a fundamentally different purpose: as the first stage in our two-stage estimation procedure for quantifying the cross-sectional relationship between volatility shocks ($\lambda_t z_t$) and return shocks.  Our empirical findings
show strong evidence of fat-tails, and we are able to use this conclusion to derive a sharpened estimate of time-varying volatility that can be used in our multi-factor model.

Another related work is that of \citet{aitsahalia:etal:2010}, who describe cross-country returns using jump models based on the class of mutually exciting Hawkes processes.  These generalizations of Poisson processes allow for a shock in one region of the world to increase the probability that future large shocks will be observed, both in the same region and in other regions.  Market shocks are modeled as instantaneous jumps in the price path, rather than as conditionally Gaussian innovations with time-varying volatility.

In the Hawkes-process model, the analogue of volatility is a latent vector of conditional Poisson jump intensities for each asset.  These intensities are assumed to follow an vector-autoregressive-like process.  Our model differs from this approach in two main ways: we model returns using stochastic volatility rather than a jump process; and we explicitly model the manner in which volatility shocks enter the cross-section of expected returns.  
By contrast, in a Hawkes-process model, there is no structural mechanism for changes in the conditional jump intensities (the analogue of volatility shocks) 
to affect the pattern of signs that one is likely to observe in the return residuals.\footnote{Regarding the issue of price-path discontinuities: 
when one marginalizes over the local scale factors $\lambda_t$ in our volatility process, one gets the increments of subordinated Brownian motion for $\sigma_t$, which will also have discontinuous sample paths as the discretization becomes arbitrarily fine.  Mathematically, the difference is between a finite-activity Poisson process superimposed on a diffusion process, and an infinite activity L\'evy process with a countable number of absolutely summable tiny jumps.  For a formal discussion of this point, see \citet{Polson:Scott:2010b}.}  There are also multivariate time-varying correlation GARCH in the mean models, such as the BEKK(1,1) model. We prefer to examine conditional volatility in a multivariate factor-model setting, much the same way as \citet{ng:engle:rothschild:1992}.  The main advantage then of our approach is being able to also account for directional volatility moves in returns, which we demonstrate in the next section to be crucial for quantifying contagion.

\subsection{Benchmarking of mean/volatility models for the U.S.~market}

We have described a model wherein heavy-tailed shocks to volatility may enter the cross-section of expected returns.  We now present evidence that the severity of the shocks that arise during financial crises cannot be explained easily either in the Gaussian SV framework, or in the asymmetric GARCH-in-mean (GARCH-AM) framework.  The explosive SV model, on the other hand, does quite well.

\begin{figure}
\begin{center}
\includegraphics[width=4.5in]{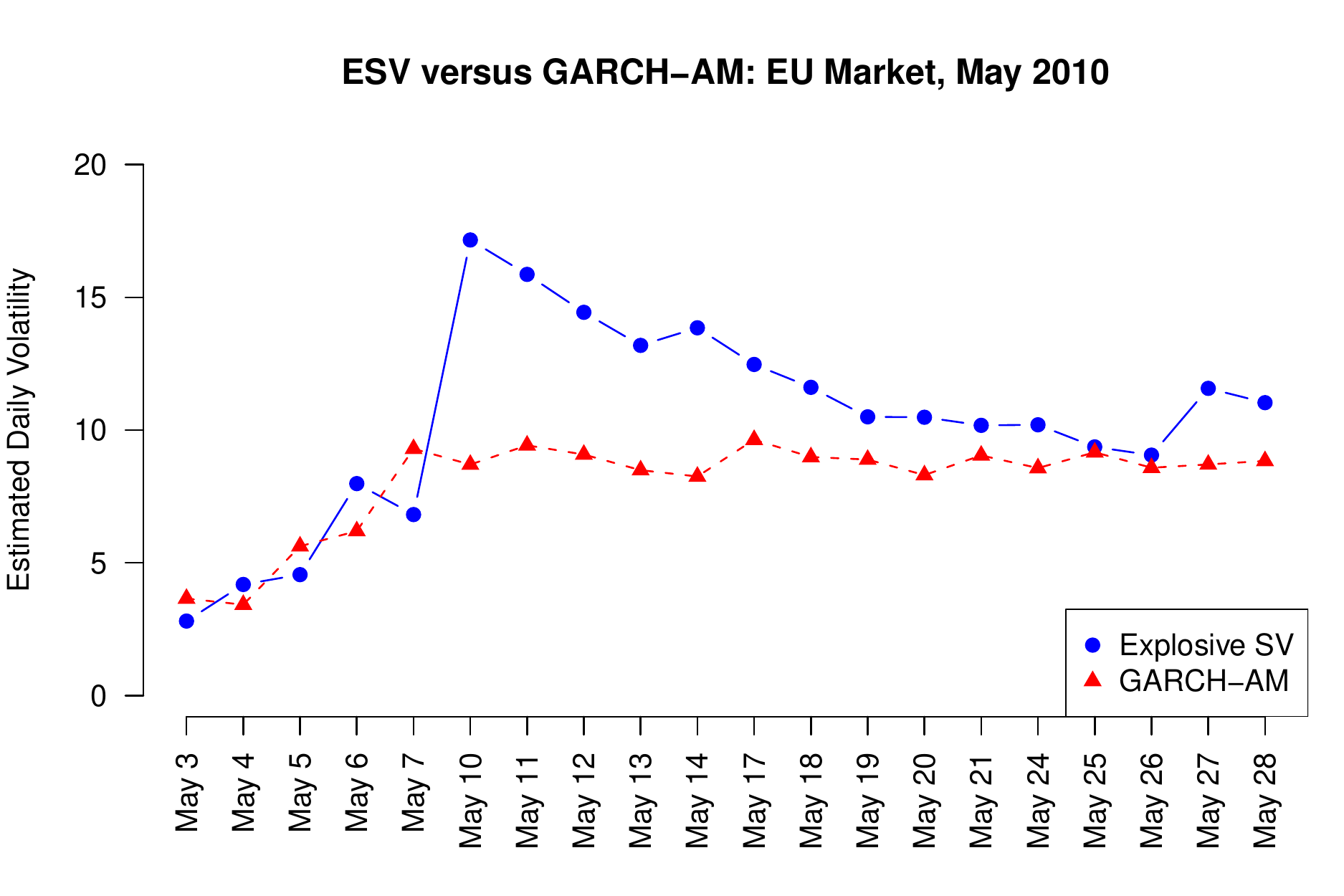}
\end{center}
\caption{\label{fig:euvolmay2010} Two estimates of daily volatility for the EU market during May 2010.}
\end{figure}

\begin{table}
\begin{center}
\begin{scriptsize}
\caption{\label{tab:biggestmoves}The ten largest residuals from GARCH-AM model fit to the U.S.~market returns 1963--2010 (top) and E.U.~market returns 2000--2010 (bottom), compared with the corresponding residuals from the ESV model.  The model residuals are standardized by the estimated conditional volatility (thus 1 denotes a one-standard-deviation move).}
\vspace{1pc}
\begin{tabular}{r r c rr}
		&				&& \multicolumn{2}{c}{Model Residual} \\
Date		& US Return (\%)		&& GARCH-AM & ESV \\
\hline
July 7, 1986 & -2.86 && -4.92 & -2.37\\
September 11, 1986 & -4.37 && -5.13 & -2.53\\
October 19, 1987 & -17.16 && -7.76 & -4.91\\
October 13, 1989 & -5.37 && -10.27 & -2.97\\
November 15, 1991 & -3.43 && -6.84 & -2.36\\
February 16, 1993 & -2.60 && -5.09 & -2.28\\
February 4, 1994 & -2.29 && -5.46 & -2.20\\
October 27, 1997 & -6.55 && -6.55 & -2.07\\
January 4, 2000 & -3.99 && -6.05 & -2.37\\
February 27, 2007 & -3.43 && -7.34 & -2.35 \\
\\
		&				&& \multicolumn{2}{c}{Model Residual} \\
Date		& EU Return (\%)		&& GARCH-AM & ESV \\
\hline
January 4, 2001 & -4.08 && -3.56 & -2.32\\
January 3, 2006 & 3.52 && 4.41 & 2.38\\
May 17, 2006 & -3.63 && -3.37 & -2.33\\
February 27, 2007 & -5.60 && -6.72 & -2.70\\
July 26, 2007 & -4.25 && -3.48 & -2.53\\
September 4, 2008 & -5.23 && -4.19 & -2.76\\
September 29, 2008 & -11.11 && -4.86 & -3.62\\
April 27, 2010 & -5.18 && -4.12 & -2.69\\
August 11, 2010 & -5.46 && -3.75 & -2.50\\
October 19, 2010 & -5.37 && -4.43 & -2.61
\end{tabular}
\end{scriptsize}
\end{center}
\end{table}

\begin{table}
\begin{scriptsize}
\begin{center}
\caption{\label{tab:marketmodelresults}Results from the U.S.~market model regressions.}
\vspace{1pc}
\begin{tabular}{ll c r c r c rr c r}
\multicolumn{2}{c}{Model}  && \multicolumn{1}{c}{$\alpha_0 \times 100$} && \multicolumn{1}{c}{$\alpha_1 \times 100$} &&  \multicolumn{2}{c}{Residuals} && \\
Mean & Volatility && Est. [$t$-stat] && Est. [$t$-stat] && Skewness ($p$) & Ex.~Kurt.~($p$) && Log-like \\
\hline
\\  
1(a)	& GARCH-AM && 3.0 [3.686] && -0.2 [-0.190] && $-0.42$ ($0.000$) & $2.62$ ($0.000$) && $-14029.97$ \\  
	& SV 		&& 9.8 [9.630] && -8.6 [-6.139] && $-0.10$ ($0.003$) & $1.34$ ($0.000$) && $-12927.04$ \\ 
	& ESV 		&& 8.7 [9.811] && -6.8 [-5.693] && $-0.02$ ($0.565$) & $-0.01$ ($0.806$) && $-12833.40$ \\ 
\\
2(a)	& GARCH-AM && 3.2 [1.802] && -0.4 [-0.159] && $-0.42$ ($0.000$) & $2.63$ ($0.000$) && $-14029.97$ \\  
	& SV 		&& 16.4 [8.188] && -16.5 [-6.303] && $-0.11$ ($0.001$) & $1.13$ ($0.000$) && $-12935.19$ \\ 
	& ESV 		&& 14.8 [8.365] && -14.5 [-6.168] &&  $-0.06$ ($0.088$)  & $0.11$ ($0.015$) && $-12853.19$ \\ 
\\
1(b)	& GARCH-AM && \multicolumn{1}{c}{---} && 2.4 [2.861] && $-0.42$ ($0.000$) & $2.60$ ($0.000$) && $ -14037.04$ \\  
	%& SV && --- && 0.0142 [1.0884] && $< 10^{-15}$ & $< 10^{-15}$ && -10892.41 \\   
	%& ESV && --- && 0.0142 [1.0884] && $< 10^{-15}$ & $< 10^{-15}$ && -10892.41 \\
\\
2(b)	& GARCH-AM && \multicolumn{1}{c}{---} && 5.6 [6.119] && $-0.41$ ($0.000$) & $2.59$ ($0.000$) && $ -14033.45$ \\  
	%& SV && --- && 0.0142 [1.0884] && $< 10^{-15}$ & $< 10^{-15}$ && -10892.41 \\   
	%& ESV && --- && 0.0142 [1.0884] && $< 10^{-15}$ & $< 10^{-15}$ && -10892.41 \\ 	
\\
3	& GARCH-AM && 2.9 [4.703] && \multicolumn{1}{c}{---} && $-0.42$ ($0.000$) & $2.62$ ($0.000$) && $-14029.98$ \\  
	%& SV && --- && 0.0142 [1.0884] && $< 10^{-15}$ & $< 10^{-15}$ && -10892.41 \\   
	%& ESV && --- && 0.0142 [1.0884] && $< 10^{-15}$ & $< 10^{-15}$ && -10892.41 \\ 	
\end{tabular}
\end{center}
\end{scriptsize}
\end{table}

The lag-1 asymmetric GARCH model  \citep{glosten:jag:runkle:1993} holds that today's volatility is a deterministic function of previous market shocks and volatility states.  Moreover, volatiliy is assumed to respond asymmetrically to positive and negative shocks:
$$
\sigma^2_t = \zeta_0 + \zeta_1 \epsilon_{t-1}^2 + \zeta_2  \epsilon_{t-1}^2 (\mathbf{1}_{\epsilon_{t-1} < 0}) + \zeta_3 \sigma^2_{t-1} \, ,
$$
inducing returns to exhibit the so-called ``leverage effect.''  When volatility appears in the model for expected returns, as here, we have an asymmetric GARCH-in-mean model (GARCH-AM).

We fit these five mean models cross with the three volatility models (GARCH-AM, Gaussian SV, ESV) to the daily excess returns on the European and US market portfolios, considered independently.  The maximum-likelihood estimate of the GARCH model was computed using gradient descent.  To fit the SV models, we used a sequential Monte Carlo algorithm called particle learning (PL).  For an explanation of the particle-learning algorithm, see \citet{carvalho:etal:2010}, along with \citet{lopes:tsay:2011}.  Each SV model has three unknown parameters associated with the volatility process: the long-term mean of volatility $(\sigma_0)$, the persistence of volatility ($\phi$), and the scale of the volatility shocks ($\tau^2$).  In traditional empirical-Bayes fashion, we searched a grid of these values to find the combination that maximized the log-likelihood of the data for different model considered.  For the ESV model, we fixed the degrees-of-freedom parameter to $\nu = 2$.

The data set used for estimating the U.S.~market model runs from July 1, 1963 to October 29, 2010; comprises $11916$ daily returns; and is freely available from Kenneth French's website.\footnote{\url{http://mba.tuck.dartmouth.edu/pages/faculty/ken.french/Data_Library/f-f_factors.html}}  The excess return on the US market is the value-weighted return on all NYSE, AMEX, and NASDAQ stocks, minus the daily return on a one-month Treasury bill. Likewise, the excess return on the Euro-wide market is the return on the Vanguard ETF corresponding to the pan-European index from Morgan Stanley Capital International, minus the daily return on a one-month Treasury bill.  The data set used for estimation runs from March 11, 2005 to October 29, 2011; and is freely available from Yahoo!~Finance, using the ticker symbol VGK.

Table \ref{tab:biggestmoves} and Figure \ref{fig:euvolmay2010} help to convey a sense of the differences that can arise between the ESV model and the asymmetric GARCH-in-mean model (GARCH-AM), another state-of-the-art volatility estimator.  (See Section \ref{sec:volatility} for details.)  Both of these models produce an expected value $\mu_t$ and an estimated standard deviation $\sigma_t$ for the daily excess return, denoted $x_t$.  But these estimates for the state variables can be strikingly different.  For example, Figure \ref{fig:euvolmay2010} plots the estimates of daily volatility ($\sigma^2_t$) from both models for May of 2010, when the Greek sovereign-debt crisis reached a climax and the assessment of the two models temporarily diverged.  The main difference arose on the 10th of May, when EU finance ministers announced the creation of a 500-billion-euro loan package intended to stop Greece's debt troubles from spreading to other Euro-zone economies. The E.U.~stock market promptly rose by $9.5\%$, the fourth-largest daily gain since the adoption of the euro.  In response, the ESV estimate of volatility nearly tripled, and remained high for several days; over the same period, the GARCH-AM estimate remained roughly constant.

Neither behavior, of course, is automatically more sensible for the particular events in question.  But the residuals $(x_t - \mu_t)/\sigma_t$ provide a natural way to quantify the fit of the two models across an entire time series.  By this standard, the ESV model behaves much more sensibly.  Table \ref{tab:biggestmoves}  lists the 10 largest residuals for the GARCH-AM model for each of two data sets: the excess returns on the U.S.~market portfolio from July 1963 to October 2010, and the excess returns on the E.U.~market portfolio from August 2000 to October 2010.  If returns are conditionally Gaussian given $\mu_t$ and $\sigma_t$, there is a $50\%$ chance of seeing a single move of five (or more) standard deviations in a period of 1000 years.  If we believe the GARCH-AM estimates for $\sigma_t$, the U.S.~market saw 9 such moves in 47 years.  As the table also shows, these daily residuals are still large under the ESV model, but not absurdly so.

This snapshot communicates the essence of the argument: when markets undergo drastic shocks, our estimate of volatility must change rapidly, especially if we hope to quantify the role of volatility risk in the cross-section of daily returns.

Table \ref{tab:marketmodelresults} shows the results of a more formal model assessment, which reveals that the explosive stochastic volatility model, version 1(a), has the largest log-likelihood.  (Models 1b, 2b, and 3 were not competitive; we include their GARCH estimates for comparison.)  We also use two other tests of model fit that assess the third and fourth moments of the normalized residuals, which should correspond to those of a standard $\N(0,1)$ distribution. First, we applied D'Agostino's test for skewness, where the null hypothesis is that residuals have zero skewness.  Second, we applied Anscombe's test, where the null hypothesis is that the residuals have no excess kurtosis, to detect tails that are systematically different from the standard normal distribution.  Table \ref{tab:marketmodelresults} also reports the empirical skewness and kurtosis of the standardized residuals for each of the models, along with the two-sided $p$-values for D'Agostino's test and Anscombe's test.  Both of these tests strongly contra-indicate the GARCH-AM and Gaussian SV models, but not the explosive stochastic-volatility model.

\section{Quantifying contagion in the EU sovereign debt crisis}
\label{sec:contagion-empirical}

Using Model 1(a) in conjuction with an explosive stochastic volatility model, we estimated filtered means and volatility states for both the US and EU market risk factors.  We used these results to construct the predictors in the four-factor model implied by Equation (\ref{eqn:conditionalmodel2}). The first two factors are returns on market-baskets, while the other two are our correspond to our filtered explosive-volatility factors:
$$
E(y_{it}| EU , US ) = \beta_i^{US} x^{US}_{t} + \beta_i^{EU} x^{EU}_{t} + \gamma_i^{US} \delta^{US}_{t} + \gamma_i^{EU} \delta^{EU}_{t} \, ,
$$
where $y_{it}$ is the return on index $i$, $\delta^{US}_{t}$ is the volatility shock to the US market, and $\delta^{EU}_{t}$ is the excess volatility shock to the European market.  The excess shock is defined as the residual after regressing the EU volatility shock upon the US volatility shock.  This measures the part of the EU volatility shock that cannot be explained by shared dependence; this is necessary to avoid marked collinearity, since the US volatility shock strongly predicts the EU volatility shock.   In constructing our volatility risk factors, we use filtered volatility estimates, 
rather than smoothed estimates (which investors could not, in principle, know contemporaneously).

We first fit this model assuming static values of all factor loadings.  These parameters are crucial in that they measure the market co-movements that can be explained by cross-section dependence upon volatility shocks.  Recall, moreover, from Section 3 that there is a direct correspondence between the $\gamma$ parameters and the covariance matrix between return and volatility shocks.  By estimating $\gamma$, we are reconstructing this covariance matrix.

Our estimates of these parameters are summarized in Table \ref{tab:contagioneffects1}.  (In this table and others, ``Euro'' refers to the daily return on an exchange-traded funded that holds Euros in cash.)  These results demonstrate excess correlation in the residuals after controlling for the first two factors (that is, relative to asset-pricing model that includes global and regional market risk).  Notice the stark difference in magnitude of the $\gamma_i$ coefficients when the ESV model is used to estimate factor volatilities, as opposed to the GARCH-AM model.  Our findings suggest that there is a real, substantively meaningful effect of volatility shocks in the cross-section of returns---but that this effect is subtle, and can easily be masked unless one pays very careful attention to estimating volatility states themselves.  A comparison of the magnitude of the coefficient estimates is particularly instructive here.  Even though the GARCH-AM volatility model leads to many statistically significant loadings, the practical significance of these effects is largely muted, often by an order of magnitude or more.

\begin{figure}
\begin{center}
\includegraphics[width=6.0in]{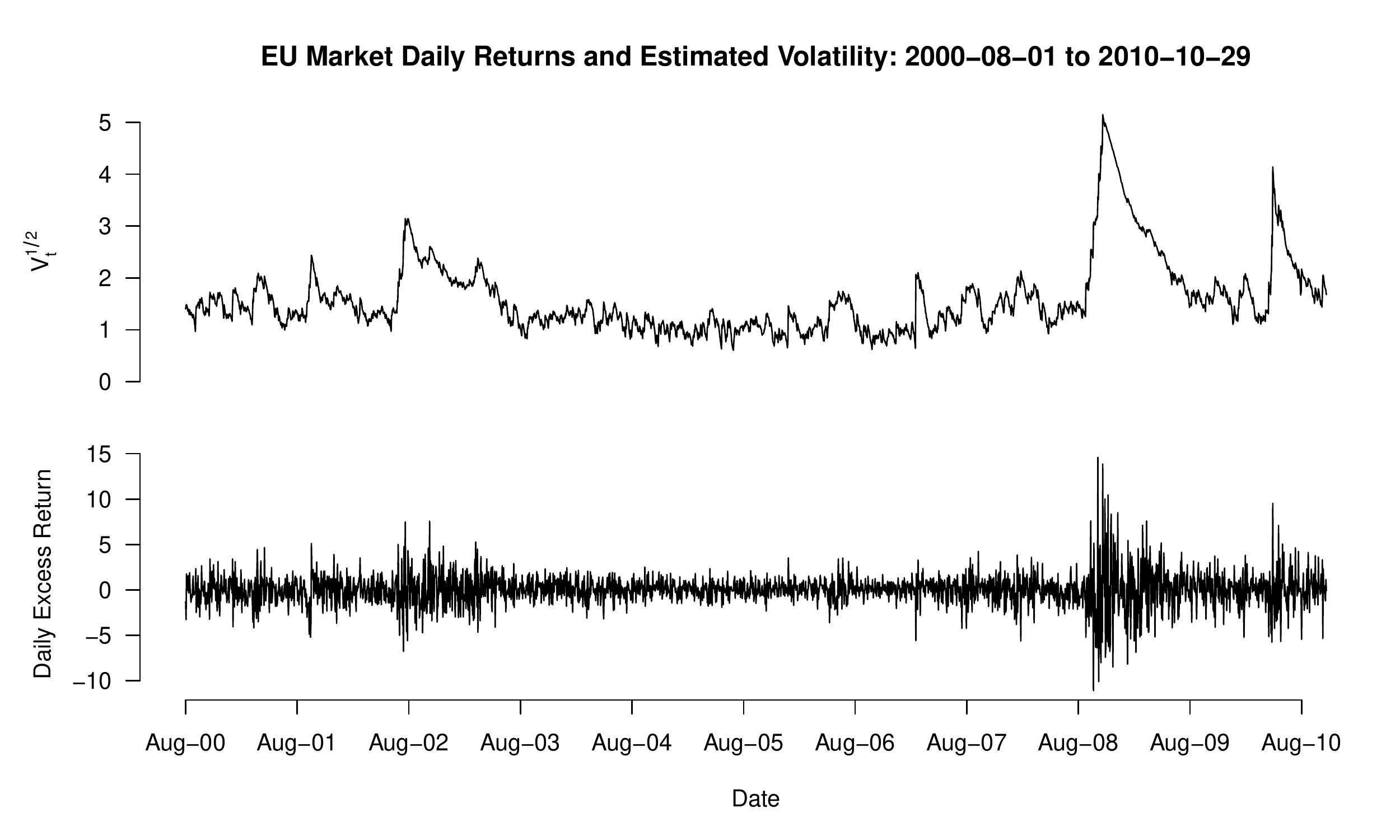}
\end{center}
\end{figure}

\begin{figure}
\begin{center}
\includegraphics[width=6.0in]{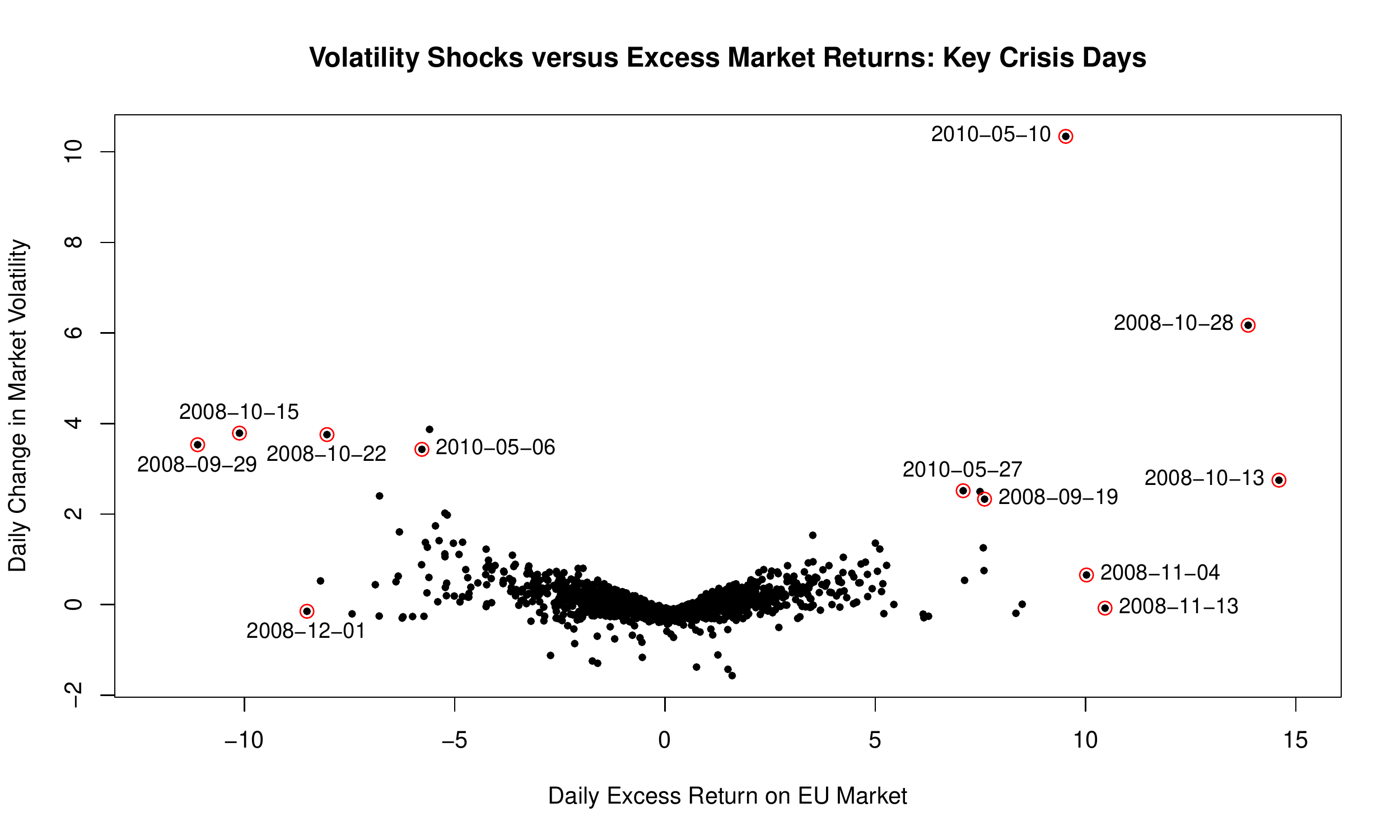}
\end{center}
\end{figure}

\begin{table}
\begin{center}
\begin{footnotesize}
\caption{\label{tab:contagioneffects1}The results from fitting the four-factor model incorporating contagion effects due to differential transmission of explosive volatility shocks.  Loading estimates are listed with their t-statistics in brackets.  Countries are listed in ascending order of their loadings on the EU residual volatility factor.  We include estimates of model parameters under both the GARCH-AM and ESV models, to demonstrate how our conclusions regarding the importance of volatility in the cross-section only emerge under the superior volatility model (as measured by log-likelihood from the previous section).  ``Euro'' refers to the daily return on an exchange-traded funded that holds Euros in cash.}
\vspace{1pc}
\begin{tabular}{r c  rr  c  rr  c  rr  c  rr}
Explosive SV && \multicolumn{2}{c}{$\beta_i^{US}$} && \multicolumn{2}{c}{$\beta_i^{EU}$} && \multicolumn{2}{c}{$\gamma_i^{US}$} && \multicolumn{2}{c}{$\gamma_i^{EU}$} \\
\midrule
DEU  &  & $0.125$ & [\phantom{$-$}$4.30$] && $0.903$ & [$38.16$] && $0.444$ & [\phantom{$-$}$5.44$] && $-0.229$ & [$-2.79$]  \\
Euro  && $-0.443$ & [$-19.7$] && $0.524$ & [$28.57$] && $0.050$ & [\phantom{$-$}$0.78$] && $-0.17$ & [$-2.68$] \\
UK  && $0.200$ & [\phantom{$-$}$6.80$] && $0.810$ & [$33.91$] && $0.141$ & [\phantom{$-$}$1.71$] && $-0.180$ & [$-2.18$]  \\
SWE  && $0.177$ & [\phantom{$-$}$3.88$] && $1.084$ & [$29.22$] && $-0.247$ & [$-1.93$] && $-0.221$ & [$-1.71$]  \\
BEL  && $0.119$ & [\phantom{$-$}$2.95$] && $0.824$ & [$25.10$] && $-0.290$ & [$-2.56$] && $-0.162$ & [$-1.42$]  \\
\\
CHF  && $0.057$ & [\phantom{$-$}$1.93$] && $0.719$ & [$29.78$] && $-0.186$ & [$-2.23$] && $0.011$ & [\phantom{$-$}$0.13$]  \\
FRA  && $0.062$ & [\phantom{$-$}$2.92$] && $0.972$ & [$55.83$] && $0.058$ & [\phantom{$-$}$0.97$] && $0.045$ & [\phantom{$-$}$0.75$]  \\
ESP  && $0.003$ & [\phantom{$-$}$0.08$] && $1.047$ & [$37.52$] && $0.018$ & [\phantom{$-$}$0.19$] && $0.097$ & [\phantom{$-$}$1.00$] \\
NLD  && $0.132$ & [\phantom{$-$}$4.76$] && $0.879$ & [$39.00$] && $-0.236$ & [$-3.03$] && $0.195$ & [\phantom{$-$}$2.49$]  \\
ITA  && $0.023$ & [\phantom{$-$}$0.70$] && $0.994$ & [$37.48$] && $-0.006$ & [$-0.07$] && $0.25$ & [\phantom{$-$}$2.72$] \\
\\
GARCH-AM && \multicolumn{2}{c}{$\beta_i^{US}$} && \multicolumn{2}{c}{$\beta_i^{EU}$} && \multicolumn{2}{c}{$\gamma_i^{US}$} && \multicolumn{2}{c}{$\gamma_i^{EU}$} \\
\midrule
DEU  && $0.1$ & [$4.1$] && $0.92$ & [$44.4$] && $0$ & [$-1.6$] && $0$ & [$-3$]  \\
Euro  && $-0.44$ & [$-21.8$] && $0.52$ & [$32.6$] && $-0.01$ & [$-3.1$] && $-0.02$ & [$-2.5$]  \\
UK  && $0.23$ & [$8.8$] && $0.77$ & [$37.2$] && $0$ & [$-1.1$] && $0$ & [$-0.9$] \\
SWE  && $0.19$ & [$4.8$] && $1.06$ & [$32.7$] && $0.02$ & [$3.6$] && $-0.01$ & [$-1.5$]  \\
BEL  && $0.09$ & [$2.5$] && $0.84$ & [$30.4$] && $-0.01$ & [$-1.3$] && $-0.02$ & [$2.1$]  \\
\\
CHF  && $0.06$ & [$2$] && $0.71$ & [$33$] && $0.01$ & [$2.2$] && $-0.01$ & [$-1$]  \\
FRA  && $0.04$ & [$2$] && $1$ & [$65.4$] && $0$ & [$1.5$] && $-0.03$ & [$-5.9$]  \\
ESP  && $-0.07$ & [$-2.4$] && $1.11$ & [$42.7$] && $0.01$ & [$2.4$] && $-0.01$ & [$-0.8$]  \\
NLD  && $0.13$ & [$4.9$] && $0.88$ & [$44.3$] && $0$ & [$0$] && $-0.01$ & [$-0.8$]  \\
ITA  && $-0.05$ & [$-1.5$] && $1.07$ & [$44.2$] && $-0.01$ & [$-2$] && $-0.02$ & [$-4$] 
\end{tabular}
\end{footnotesize}
\end{center}
\end{table}

\begin{sidewaystable}
\begin{center}
\begin{scriptsize}
\caption{\label{tab:contagioneffects2}The results from fitting the four-factor model incorporating contagion effects due to differential transmission of explosive volatility shocks, with different loadings on the regional excess volatility factor during crisis times UP THROUGH 4 OCT 2011.  Loading estimates are listed with their t-statistics in brackets.}
\vspace{1pc}
\begin{tabular}{r  rr  c  rr  c  rr  c  rr c rr c rr c rr c c}
Country & \multicolumn{2}{c}{$\beta_i^{US}$} && \multicolumn{2}{c}{$\beta_i^{EU}$} && \multicolumn{2}{c}{$\gamma_i^{US}$} && \multicolumn{2}{c}{$\gamma_i^{EU}$, no crisis} && \multicolumn{2}{c}{$\gamma_i^{EU}$, Sep/Oct '08} && \multicolumn{2}{c}{$\gamma_i^{EU}$, May '10} && \multicolumn{2}{c}{$\gamma_i^{EU}$, Aug '11} && $p_{\gamma}$ \\
\midrule
Euro  & $-0.44$ & [$-21.5$] && $0.52$ & [$32.2$] && $0.04$ & [$1.2$] && $-0.08$ & [$-1.4$] && $0.01$ & [$0.1$] && $-0.6$ & [$-4.6$] && $-0.03$ & [$-0.3$] && 0.000 \\
DEU  & $0.11$ & [$4.2$] && $0.93$ & [$44.2$] && $0.19$ & [$4$] && $0.02$ & [$0.3$] && $-0.29$ & [$-2$] && $-0.76$ & [$-4.6$] && $-0.41$ & [$-2.8$] && 0.000 \\
UK  & $0.24$ & [$9.4$] && $0.76$ & [$37$] && $0.08$ & [$1.9$] && $0.11$ & [$1.5$] && $-0.84$ & [$-5.8$] && $-0.31$ & [$-1.9$] && $-0.17$ & [$-1.2$] && 0.000 \\
\\
FRA  & $0.04$ & [$1.9$] && $1$ & [$64$] && $0.04$ & [$1.1$] && $0.1$ & [$1.8$] && $0.21$ & [$1.9$] && $-0.06$ & [$-0.5$] && $-0.39$ & [$-3.6$] && 0.000 \\
CHF  & $0.05$ & [$1.8$] && $0.72$ & [$33$] && $-0.09$ & [$-1.9$] && $-0.1$ & [$-1.3$] && $0.35$ & [$2.3$] && $-0.17$ & [$-1$] && $0.42$ & [$2.8$] && 0.001 \\
SWE  & $0.18$ & [$4.5$] && $1.07$ & [$32.8$] && $-0.15$ & [$-2.1$] && $0.03$ & [$0.2$] && $0.13$ & [$0.6$] && $-0.04$ & [$-0.2$] && $-0.04$ & [$-0.2$] && 0.852 \\
NLD  & $0.12$ & [$4.9$] && $0.88$ & [$43.5$] && $-0.06$ & [$-1.5$] && $-0.02$ & [$-0.3$] && $0.16$ & [$1.1$] && $0.19$ & [$1.2$] && $-0.09$ & [$-0.6$] && 0.164  \\
\\
ITA  & $-0.05$ & [$-1.7$] && $1.07$ & [$44.1$] && $0.01$ & [$0.1$] && $-0.12$ & [$-1.4$] && $0.5$ & [$2.9$] && $0.46$ & [$2.4$] && $-0.33$ & [$-1.9$] && 0.000 \\
BEL  & $0.1$ & [$2.7$] && $0.82$ & [$29.6$] && $-0.26$ & [$-4.3$] && $-0.31$ & [$-3.2$] && $0.07$ & [$0.3$] && $0.68$ & [$3.1$] && $0.38$ & [$1.9$] && 0.000 \\
ESP  & $-0.08$ & [$-2.3$] && $1.11$ & [$42.7$] && $0.01$ & [$0.3$] && $-0.15$ & [$-1.7$] && $0.35$ & [$1.9$] && $0.89$ & [$4.3$] && $-0.38$ & [$-2.1$] && 0.000
\end{tabular}
\end{scriptsize}
\end{center}
\end{sidewaystable}

The main empirical utility of our model, however, is in studying the time-varying behaviour of the volatility loadings $ \gamma_i^{US} , \gamma_i^{EU} $, particularly during crisis periods. To explore this effect, we introduce use dummy variables to measure period-by-period variation in the loadings on the volatility shocks.   We include dummies for the two most notable ``EU-centric'' crisis periods of May 2010 and August 2011.

For comparative purposes, we will also include the loading estimates for the financial crisis of 2008.
Late summer and fall of that year provided the height of the financial crisis. 
On Monday, September 15, 2008 the Dow Jones plunged 504 points with the US authorities trying to put a rescue
package together for AIG for a \$20 billion lifeline. On Tuesday September 30, Asian markets were the first to react to the shock that the \$700 billion
Wall Street bailout had failed to pass through Congress.  On Wednesday, October $15$, 2008, the FTSE suffered its fifth biggest fall in history, closing down $7.16$\% at $4079.5$ (a $315$ point fall); the Dow Jones dropped by $7.8$\%. The following day, the self-exciting nature of these shocks led (on Thursday, October $16$) to Japan's Nikkei suffering its worst fall since 1987, and to the FTSE100 slumping again by a further 218 points (to 3861).

These events form a useful ``control population,'' in that they are global rather than EU-centric, and thus we would expect to see different patterns of loadings on the volatility shocks during these two periods.  To quantify the need for dynamic structure, we have also provided a $p$-value for the partial $F$-test corresponding to a test of the time-varying contagion effect model, versus no time-varying effect.

The results from our basic pricing model are provided in Table \ref{tab:contagioneffects1}.  Clearly different countries load upon aggregate volatility shocks in different, economically interpretable ways during different crises; thus the factor loadings cannot be explained simply by nonlinear dependence of returns on volatility.   For example, Germany has significant negative loadings on volatility shocks during the major crisis periods of May 2010 and August 2011.  These two contagion terms have individually significant $t$-statistics, and the $F$-test strongly supports their inclusion in the model.

An interesting comparison is with Spain and Italy.  These two indices have significantly significant positive loadings on volatility shocks during 2010, and significantly negative loadings during August 2011.  In the May 2010 period, Germany had to play a large role in bailing out Greece, raising the probability of a bailout for other debt-stricken nations.  But in August 2011, the prospects for a satisfactory resolution to the Greek situation looked significantly worse, and market fears about Spain and Italy intensified.

For France, there are a number of differences between the periods of May 2010 and August 2011. In particular, the major concerns over the French banks Societe Generale, BNP Paribas, and Credit Agricole weren't present in 2010; all of these banks faced downgrades by ratings agencies in 2011, due to their massive exposure to Greek bonds.

Also instructive is the pattern of loadings exhibited by Switzerland (CHF).  During both the 2008 and 2011 crisis periods, the Swiss Franc was seen as a safe haven for investors---particularly in 2011, when fears gathered about the possible dissolution of the euro.  (The franc went from 0.77 euros in early April to 0.97 euros on August 10, 2011; no such spike was observed during May of 2010.)  The corresponding estimates during these periods suggest a noticeable ``flight to quality'' effect.  Finally, Sweden---which has been relatively unaffected by the EU debt crisis---was the only country where the volatility shocks were consistently irrelevant.

\section{Conclusion} 

The recent European debt crisis has brought to focus the inter-dependence of market movements.
The increasingly explosive behavior of returns and volatility in periods of market turmoil
has heightened the need for statistical models capable of describing contagion effects, such as those eminating from of the Greek debt crisis.  These effects are hard to measure, with the challenge arising from the facts that excess correlation can be caused by co-movements in an aggregate risk factor, and that conditional factor means and volatilities cannot be observed directly.

This leads us to consider dynamic models of the expected return/risk trade-off, a relationship that is fundamental to any asset pricing model and to models which can uncover time-varying correlations.  To uncover the effects of explosive volatility in explaining contagion, we propose a four factor model, fit to data from major European equity markets during the period 2008-2011.  In this model, loadings on the volatility risk factors have natural interpretations in terms of time-varying patterns of correlation in a mutually exciting stochastic volatility model (Section 3), whereby shocks to returns and shocks to volatility are correlated cross-sectionally.  We construct the volatility risk factors themselves using a heavy-tailed explosive stochastic volatility model, which is comprehensively benchmarked against the popular GARCH-AM framework.

Our findings are two-fold.  First, traditional volatility models are not sufficiently sensitive to extreme market movements, and are thus inefficient at estimating conditional market volatility.  Secondly,  after controlling for global and regional market integration, we still find evidence for contagion in the European debt crisis. We have gone beyond merely demonstrating the presence of excess correlation, however.  Indeed, we have shown that some of this excess correlation can be explained by postulating volatility effects in the cross section of returns, paralleling the construction of \citet{ang:hodrick:JOF:2006} for domestic equities.  But these effects only emerge when the explosive SV model is used; they are largely muted when the demonstrably inferior GARCH-AM model is used instead.  This foregrounds the need for efficient, state of the art volatility estimation techniques in studying presumptive periods of financial contagion.

One deficit of our model is that it is purely statistical nature, and is therefore silent on the underlying macroeconomic causes of such events.  However, we feel that this data period and our multi-factor approach provides a benchmark for future empirical finance studies of contagion,  particularly given the readily interpretable nature of our findings (e.g.~Table \ref{tab:contagioneffects2}).  Moreover, as the literature review in Section 2 explains, there are many theoretical reasons why the study of aggregate market volatility, and the transmission of volatility shocks, should play an important role in understanding contagion.  These theories---which encompass a myriad of potential causes, from behavioral to macroeconomic to financial---are buttressed by the fact that the explanatory power of volatility shocks in the cross section appears to be strongest during periods of crisis.

\begin{footnotesize}
\singlespace
\bibliographystyle{abbrvnat}
\bibliography{masterbib,business_articles}
\end{footnotesize}

\end{document}